\documentclass[apj]{emulateapj}
\usepackage{apjfonts}

\shorttitle{Dark matter halos of barred disk galaxies}
\shortauthors{Cervantes Sodi, Li \& Park}

\begin{document}

\title{Dark matter halos of barred disk galaxies}

\author{Bernardo Cervantes Sodi \altaffilmark{1,2}, Cheng Li \altaffilmark{3}, 
Changbom Park \altaffilmark{2} }
\altaffiltext{1}{Centro de Radioastronom\'ia y Astrof\'isica, Universidad Nacional Aut\'onoma de M\'exico, Campus Morelia, A.P. 3-72, C.P. 58089 Michoac\'an, M\'exico, \textit{b.cervantes@crya.unam.mx}}
\altaffiltext{2}{Korea Institute for Advanced Study, Dongdaemun-gu, Seoul 130-722, Republic of Korea}
\altaffiltext{3}{
Key Laboratory for Research in Galaxies and Cosmology of Chinese Academy
of Sciences, Shanghai Astronomical Observatory, Nandan Road 80, Shanghai 200030, China}

\begin{abstract}
We use a large volume-limited sample of disk galaxies drawn from the
Sloan Digital Sky Survey Data Release 7 to study the dependence of the bar fraction on
the stellar-to-halo mass ratio, making use of a group catalog, we identify central and
satellite galaxies in our sample. For the central galaxies in the sample we estimate
the stellar-to-halo mass ratio (M$_{\mathrm{*}}/$M$_{\mathrm{h}}$)
and find that the fraction of barred galaxies is a strong
function of this ratio, especially for the case of strong bars. Bars are
more common in galaxies with high M$_{\mathrm{*}}/$M$_{\mathrm{h}}$ values,
as expected from early theoretical works that showed that systems with massive dark
matter halos are more stable against bar instabilities.
We find that the change of the bar fraction with M$_{\mathrm{h}}$ and M$_{\mathrm{*}}$
is stronger if we consider a relation with the form $f_{\mathrm{bar}}=f_{\mathrm{bar}}$(M$_{\mathrm{*}}^{\alpha}/$M$_{\mathrm{h}}$)
with $\alpha=1.5$, and that the bar fraction is largely independent of other physical properties
such as color and spin parameter
when M$_{\mathrm{*}}^{3/2}/$M$_{\mathrm{h}}$ is fixed.
With our sample of galaxies segregated
into centrals and satellites, we also compare the fraction of barred galaxies in each group,
finding a slightly higher bar fraction for satellites when compared with centrals at fixed stellar mass,
but at fixed color this difference becomes very weak. This result, in agreement with previous studies, confirms that the bar fraction does not
directly depend on the group/cluster environment, but the dependence exists through
its dependence on internal morphology.

\end{abstract}


\section{Introduction}

A substantial percentage of disk galaxies at low redshift is known to
host stellar bars (e.g. de Vaucouleurs et al. 1991;
Eskridge et al. 2000; Nair \& Abraham 2010, Lee et al. 2012,
henceforth Lee+12). These prominent non-axisymmetric
structures are believed to have an important influence on
galaxy evolution, such as redistributing mass and angular
momentum between the constituents of the galaxies
(Friedli \& Benz 1993; Debattista \& Sellwood 2000;
Athanassoula 2002; Martinez-Valpuesta et al. 2006),
promoting the formation of spiral arms and rings
(Lindblad 1960; Toomre 1969; Sanders \& Huntley 1976; Schwarz 1981),
fueling gas to the centers of the galaxies (Shlosman et al. 1989;
Friedli \& Benz 1993), and helping in the build-up of pseudo-bulges
(Sheth et al. 2005; Laurikainen et al. 2007; Okamoto 2013).

Just as bars act as major agents of secular evolution
of their hosting galaxies (Athanassoula 2013; Cheung et al. 2013; Sellwood 2014),
the internal physical characteristics of their hosting galaxies play
a major role in determining the presence and evolution of bars. 
Using extensive samples of galaxies, at both low and at high redshifts,
a number of important studies have examined the dependence of the disk bar fraction
($f_{\mathrm{bar}}$) on different physical properties. While some early studies
( Aguerri et al. 2009; Barazza et al. 2009) found that bars are mostly located in
blue, low concentrated galaxies with low luminosities and masses, more recent
studies report the opposite. In general, the fraction of barred galaxies
is higher in luminous, more massive galaxies, than in their less massive,
fainter counterparts (Nair \& Abraham 2010; Lee+12; Masters et al.
2012; Oh et al. 2012; Wang et al. 2012). This non-monotonic behavior of the bar fraction
is also found as a function of galaxy color; with the bar fraction increasing
considerably when moving from blue to red systems (Nair \& Abraham 2010;
Masters et al. 2011; Lee+12; Oh et al. 2012). Masters et al. (2012) reported a 
lower bar fraction in gas-rich disk galaxies than gas-poor ones, and Wang et al. (2012)
found  that galaxies with strong bars present an enhanced central star formation rate 
or a star formation rate that is suppressed when compared to the mean, highlighting 
the important role that bars have in the quenching of star formation.
Finally, the bar fraction is higher and the bars are longer in early-type spirals, with
more prominent bulges and higher concentrations, than in late-type spirals
(Elmegreen \& Elmegreen 1985; Martin 1995; Erwin 2005; Laurikainen et al. 2007;
Hoyle et al. 2011)

In regards to the formation of bars in disk galaxies, the halos in which these
galaxies are embedded play a major role. According to the early
simulations by Ostriker \& Peebles (1973), a cold stellar disk will experience
bar instabilities if the ratio of total kinetic energy to the total potential energy
exceeds 0.14. The presence of a massive halo, supported by random orbital motions, helps to stabilize the disk by increasing
the potential energy, being more efficient a large halo than a centrally concentrated
one (Hohl 1976). Efstathiou et al. (1982) reached a similar conclusion studying a set
of N-body simulations and proposing a stability criterion that requires a massive halo component
to provide stability against bar formation (see also Christodoulou et al. 1995). 
This same stability criterion is reported (Yurin \& Springel 2014) to be a good
predictor for the formation of bar in the full cosmological contex.
The inclusion of other components, such as bulges, diminished the importance of the
stabilizing effect of the halo (Athanassoula \& Sellwood 1986).
However, Athanassoula (2002, 2003) found stronger bars in halo-dominated models
than in their disk-dominated counterparts when a live halo is implemented.
More recently, DeBuhr et al. (2012) studied gravitational
interactions between live stellar disks and their dark matter halos, finding the
stellar-to-halo mass ratio to be a primary
factor to follow bar formation and evolution, with systems showing stronger
stability against bar formation in lower mass disks for a given halo mass.

An interesting case of study in this regard are low surface brightness galaxies (LSB),
that are expected to be stable against bar formation due to low disk self-gravity
and high dark matter content. Numerical experiments trying to simulate the
formation of bars in such systems face difficulties due to the low disk masses and
if they are successful, the bars they generate are small and unstable, leading
to the formation of bulge-like structures (Mihos et al. 1997; Mayer \& Wadsley 2004).

Halo shape has also repercussion on the fate of stellar bars. When compared with
axisymmetric halos, triaxial halos induce early bar formation (Athanassoula et al. 2013),
but at later stages damp the growth and strengthening of the bar (Berentzen et al 2006;
Athanassoula et al. 2013). Only recently, numerical simulations have addressed the influence of rotating 
halos in the formation and growth of bars (Saha \& Naab 2013; Long et al. 2014),
finding that spinning halos promote bar formation, but their growth in size and
strength gets quenched with increasing spin, explaining why $f_{\mathrm{bar}}$ decreases
with increasing $\lambda$ (Cervantes-Sodi, et al. 2013).

Recently, there have been several works studying the environmental dependence
of barred galaxies. When the environment is characterized in terms of the distance
to the nearest neighbor, the likelihood of galaxies hosting bars shows a systematic
decrease as the distance decreases (Lee+12; Casteels et al. 2013; Lin et al. 2014),
suggesting that close encounters suppress the formation of bars and/or destroy them.
If the local environmental density is considered, most of the studies show an independence
of barred galaxies with environment (Giordano et al. 2010; Mart\'inez \& M\'uriel 2011; Lee+12;
Marinova et al. 2012). Skibba et al. (2012) introduced the use of a
mark clustering statistic in their calculation of the correlation function, which helped
them to find a positive overclustering, from projected separations of 150 kpc $h^{-1}$ to 3 Mpc $h^{-1}$,
for barred, bulge-dominated galaxies. Analyzing a particular halo ocupation model, they argue that their finding suggest that the barred
galaxies in their sample are central galaxies in low mass dark matter halos or satellite galaxies
in more massive halos, hosting galaxy groups. A result in the same line is
reported by Lin et al (2014) for the case of their early-type barred galaxies, that are
more strongly clustered on scales from a few 100 kpc up to 1 Mpc, than the unbarred early-type
galaxies.

In this paper we study the dependence of the bar fraction on the stellar-to-halo mass
ratio as well as differences on the bar fraction for central and satellite disk galaxies.
In Section 2 we describe the volume-limited sample. The main results and discussion
are presented in Section 3. Lastly we summarize our general conclusions in Section 4.
Throughout this paper, we use a cosmology with density parameter
 $\Omega_{\mathrm{m}}$ = 0.3, cosmological constant
 $\Omega_{\mathrm{\Lambda}}$ = 0.7 and Hubble constant written as
  $H_{0}  = 100h$km s$^{-1}$ Mpc$^{-1}$, with $h=70$.

\section{The galaxy sample}

The sample used in this work comes from a previous work by Lee+12.
It is a volume-limited sample complete down to a $r$-band absolute
magnitude brighter than $M_{\mathrm{r}} =$  -19.5 + 5log$\mathrm{h}$ and within the redshift
range 0.02 $\leq z \leq$ 0.05489, drawn from the Sloan Digital Sky Survey
Data Release 7 (DR7; Abazajian et al. 2009). The original catalog contains 33,391 galaxies that are
classified into early (E/S0) and late (Sa to Sd) types by segregating them in the color vs.
color gradient and concentration index planes (Park \& Choi 2005) plus an
additional visual inspection.

The identification of stellar bars is done by visual inspection of $g+r+i$
combined color images from the SDSS website using \textit{Visual Tools}. To
avoid selection biases by inclination, the sample of late-type galaxies is
limited to systems with $i$-band isophotal axis ratio $b/a>0.6$,
where $a$ and $b$ are the semi-major and semi-minor axes. 
From this requirement we restrict our study to 10,674
late-type galaxies. When a stellar
bar is identified, it is further classified as a strong bar if it is
larger than one quarter of the size of their host galaxies, or as a weak bar
otherwise.  As presented in Lee+12, our classification shows a good agreement with
the classification performed by Nair \& Abraham (2010). Among the 10,674 late-type galaxies
in our sample, 23.8\% (2542 galaxies) host strong bars and 6.5\% (698 galaxies) host
weak bars, giving a bar fraction of 30.4\%, a value in good agreement with
studies that detect bars by visual inspection, with typical values between
25\% to 36\% (Nair \& Abraham 2010; Giordano et al. 2011; Masters et al. 2011; Oh et al. 2012).
Furthermore, the bar fraction distribution of our sample shows dependencies with
stellar mass, color, and concentration index, in qualitative and quantitative good
agreement with the findings of Nair \& Abraham (2010), Masters et al. (2011), and
Oh et al. (2012).
For a more detailed description of the galaxy
catalog and comparisons of the classification
with previous studies (de Vaucouleurs et al. 1991; Nair
\& Abraham 2010), we refer the reader to Lee+12 and Park \& Lee
(2014 in preparation).

The physical properties required for this study, such as $r$-band absolute magnitude
M$_r$, stellar mass $M_\ast$ and $u-r$ color are extracted from the Korea Institute
for Advanced Study Value-Added Galaxy Catalog (Choi et al. 2010),
the New York University Value Added Galaxy Catalog (NYU-VAGC; Blanton et al. 2005)
and the MPA/JHU SDSS database (Kauffmann et al. 2003; Brinchmann et al. 2004).

To obtain estimates of the parent halo mass, we match the galaxies in our sample with
an updated version of the galaxy group catalog of Yang et al. (2007), finding
a match for 91.5\% of the galaxies in our sample without any apparent bias. The groups in this catalog are
identified by applying a 'friends of friends' halo finder algorithm to the SDSS DR7, with galaxies
redshifts in the range 0.01 $\leq z \leq$ 0.20 and masses as low as
10$^{11.5} h^{-1}$ M$_{\odot}$. The halo mass estimates in the catalog are obtained by
ranking the galaxy groups by their total stellar mass or luminosity, and assuming a
one-to-one relation between the total stellar mass/luminosity of the groups and
M$_{\mathrm{h}}$. Given that in our study we are interested in the dependence of the bar fraction
on the stellar mass to halo mass ratio, we opt for the M$_{\mathrm{h}}$ estimates by ranking
total luminosity of the galaxy groups, in order to avoid introducing an intrinsic bias, but similar
results are obtain if we adopt the ranking by stellar masses, or even by employing a
different group/cluster catalog (i.e. Tempel et al. 2014), where the virial masses are
obtain through dynamical considerations.

\section{Results and discussion}

\subsection{Central galaxies}

\begin{figure}
\label{distributions}
\centering
\begin{tabular}{c}
\includegraphics[width=0.4\textwidth]{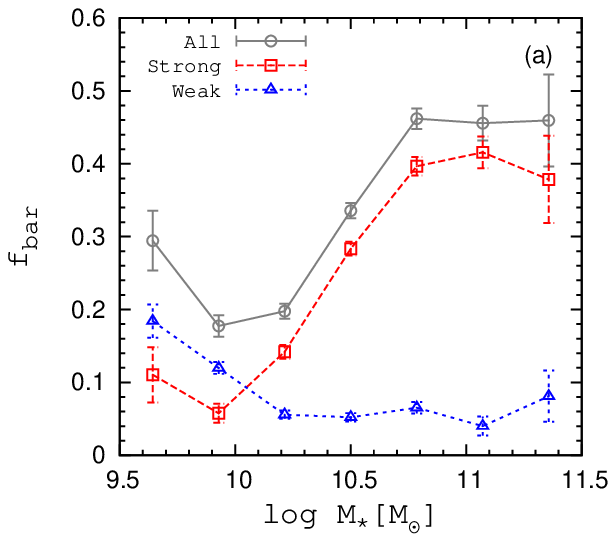} \\
\includegraphics[width=0.4\textwidth]{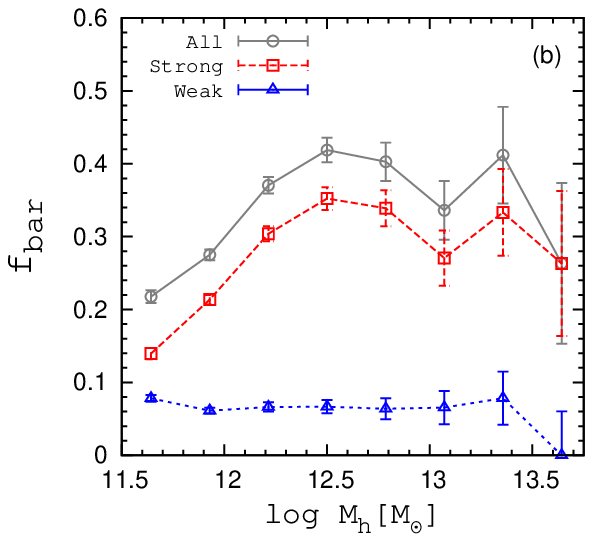} \\
\includegraphics[width=0.4\textwidth]{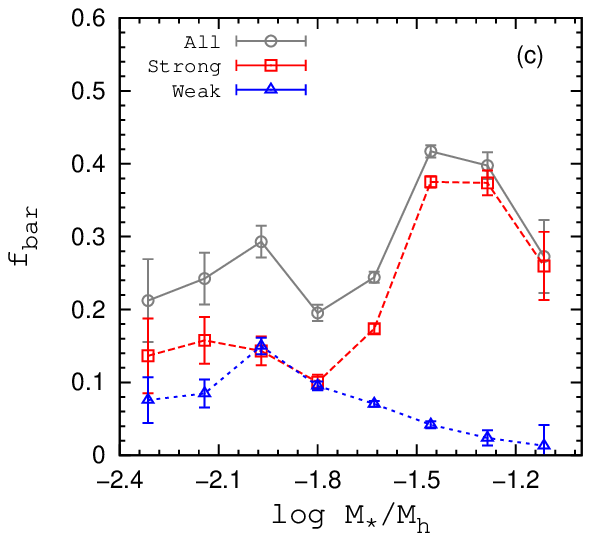} 
\end{tabular}
\caption[ ]{Fraction of barred galaxies $f_{\mathrm{bar}}$ as a function of
stellar mass M$_{\mathrm{*}}$ (\textit{a}), halo mass M$_{\mathrm{h}}$ (\textit{b}), 
and stellar-to-halo mass ratio (\textit{c}), for strong, weak, and strong plus weak bars
of our sample.
\\(A color version of this figure is available in the online journal.)
}\label{Fractions}
\end{figure}

\begin{figure*}[]
  \begin{center}
    \includegraphics [width=0.7\hsize]{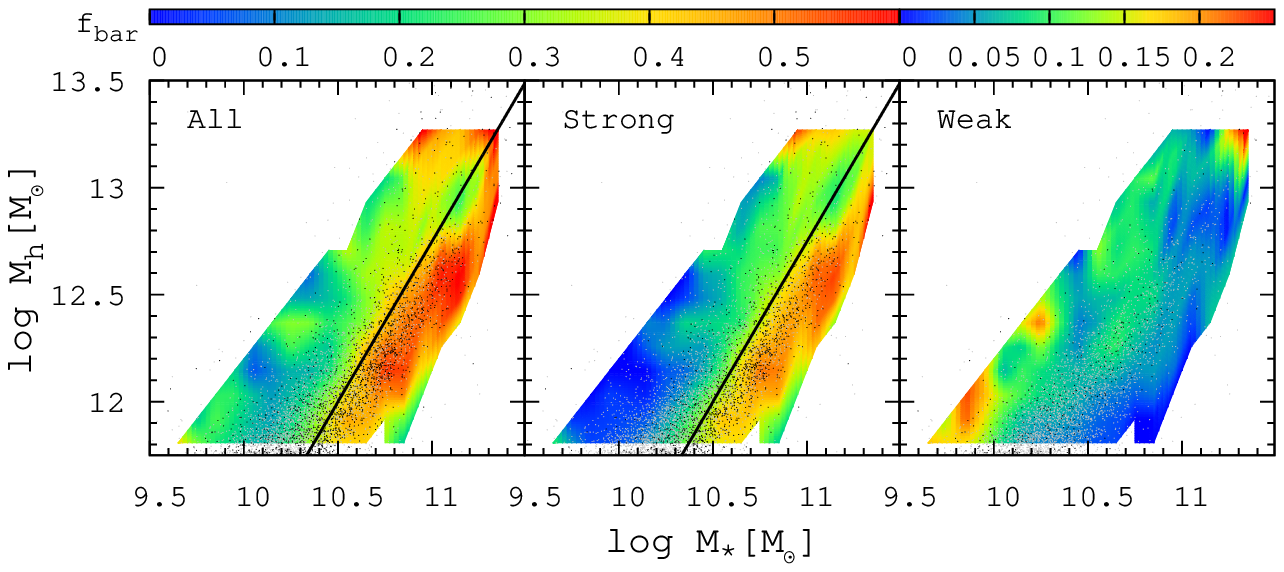}\\
    \includegraphics [width=0.7\hsize]{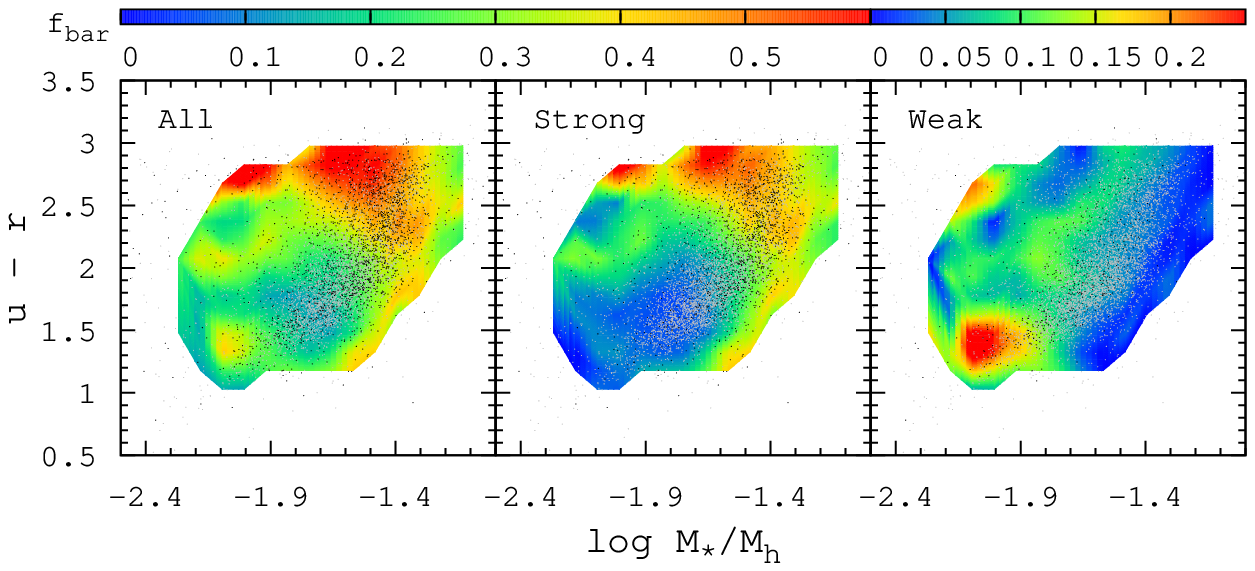} \\
    \includegraphics [width=0.7\hsize]{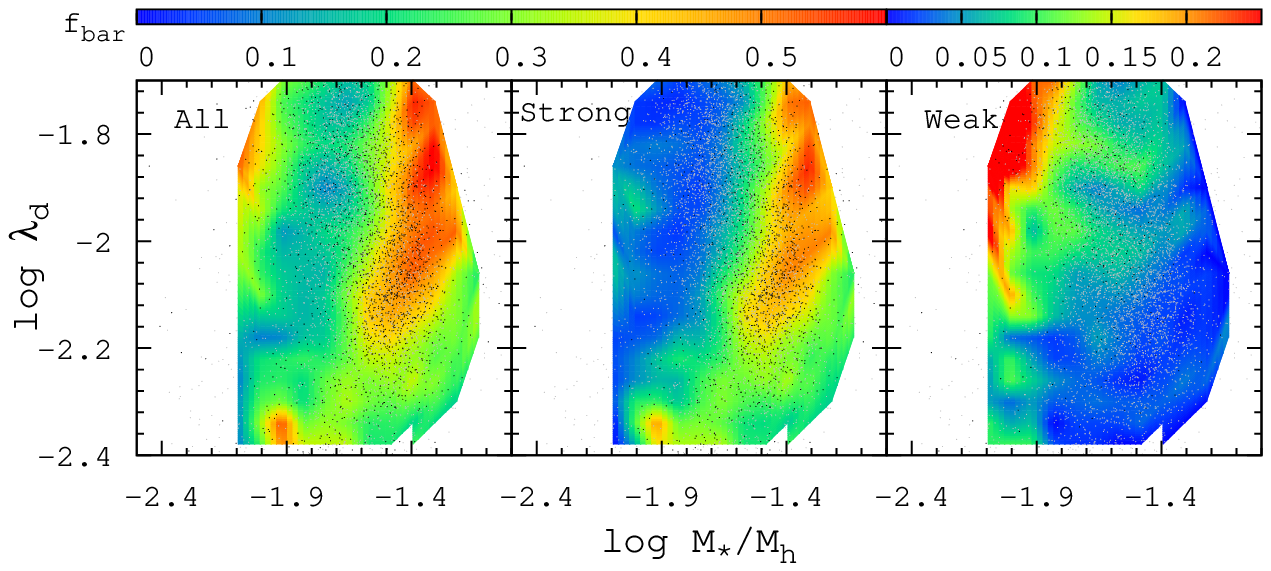}   
  
    \caption{Bar fraction $f_{\mathrm{bar}}$ isocontours in the M$_{\mathrm{h}}$ vs. M$_{\mathrm{*}}$ 
    (\textit{top panels}), $u-r$ vs. M$_{\mathrm{*}}/$M$_{\mathrm{h}}$ (\textit{middle panels}),
    and $\lambda_{d}$ vs M$_{\mathrm{*}}/$M$_{\mathrm{h}}$ (\textit{bottom panels})
    spaces. Left column correspond to strong plus
    weak bars, middle to strong bars and right to weak bars. The range and coding for $f_{bar}$ is
    the same for
    the full barred sample (strong plus weak bars) and for the limited sample of strong bars, with
    $0 \leq  f_{\mathrm{bar}} \leq 0.6$, while for the case of weak bars is restricted to
    $0 \leq  f_{\mathrm{bar}} \leq 0.25$. Gray dots represent unbarred
    galaxies, black dots represent barred ones. 
    \\(A color version of this figure is available in the online journal.)}
  \label{maps}
  \end{center}
\end{figure*}

Given that our interest is to study the dependence of the likelihood of galaxies
hosting bars on their stellar-to-halo mass ratio, in this subsection we will
include only central galaxies for which we have estimates of their respective halo
masses.

Figure~\ref{Fractions}a shows the well known dependence of $f_{\mathrm{bar}}$ on stellar mass
(M\'endez-Abreu et al. 2010; Nair \& Abraham 2010; Masters et al. 2012; Oh et al. 2012; Cervantes Sodi et al. 2013)
for the central late-type galaxies in our sample, with an increase of the bar fraction
with increasing stellar mass for the case of strong bars, while weak bars show the
opposite trend, an increase of $f_{\mathrm{bar}}$ for decreasing stellar mass. Figure~\ref{Fractions}b
 shows the dependence of $f_{\mathrm{bar}}$ on the host halo mass. Similar
to the case of stellar mass, we find that strong bars tend to be more common in
galaxies with massive halos (contrary to the results of Mart\'inez \& Muriel 2011,
who found no dependence on halo mass), while weak bars show almost no dependence of
$f_{\mathrm{bar}}$ on halo mass. Given that more massive stellar disks reside in more
massive halos, this dependence of the bar fraction on M$_{\mathrm{h}}$ is not surprising.
A more interesting feature to explore is the dependence of the bar fraction on the
stellar-to-halo mass fraction. As can be seen in Figure~\ref{Fractions}(c), for the case of strong
bars, galaxies with high M$_{\mathrm{*}}/$M$_{\mathrm{h}}$ ratios ($\geq -1.8$)
have much higher bar fraction than systems that are more dominated by dark matter. Weak bars present only a weak
dependence but in the opposite direction, with $f_{\mathrm{bar}}$ increasing with
decreasing M$_{\mathrm{*}}/$M$_{\mathrm{h}}$.

It is interesting to check if this dependence is visible at fixed stellar mass. In the top panels of Figure~\ref{maps}
 we present the bar fraction in the M$_{\mathrm{h}}$ vs. M$_{\mathrm{*}}$
plane, for the full sample of barred galaxies (strong plus weak bars, left panel), and strong
(central panel) and weak (right panel) bars separately. We use a spline kernel to
get a smooth transition of $f_{\mathrm{bar}}$, dividing the parameter space into
20 $\times$ 20 bins, and requiring at least 15 galaxies per bin to estimate the bar
fraction. The first thing to note is that
even at fixed stellar mass, there is a strong variation of $f_{\mathrm{bar}}$ with
halo mass, and the dependence is particularly clear for the case of strong bars,
with f$_{\mathrm{bar}}$ increasing with decreasing M$_{\mathrm{h}}$ at fixed 
M$_{\mathrm{*}}$, although the dependence is more dramatic fixing the halo mass
and looking at the increase of $f_{\mathrm{bar}}$ with increasing stellar mass.
Weak bars are found in galaxies with low M$_{\mathrm{*}}$ and M$_{\mathrm{h}}$ 
values.

For the cases of the full sample and the restricted subsample of strong
bars, we find that the bar fraction presents a secondary maximum at
high halo mass ($M_h \ga 10^{13}M_\odot$), but relatively low stellar
mass ($M_\ast\la 10^{11.2}M_\odot$). These systems are actually central galaxies of rich groups,
where the stellar mass refers to the central galaxy only, but the total halo mass refers
to the mass of the parent halo plus the satellite systems. In this sense, the value obtained
through our estimate for the M$_{\mathrm{*}}/$M$_{\mathrm{h}}$ represents only a lower
value or the real one. Besides, in these rich groups, other mechanisms might be taking place
changing the likelihood of galaxies hosting bars.

With the most massive galaxies populating the most massive halos and at the
same time being in general redder and with a higher M$_{\mathrm{*}}/$M$_{\mathrm{h}}$ 
than less massive galaxies, it is important to check if the dependence of
the bar fraction on M$_{\mathrm{*}}/$M$_{\mathrm{h}}$ is not only a reflection
of the dependence of M$_{\mathrm{*}}/$M$_{\mathrm{h}}$ on color. Figure~\ref{maps} middle panels show
the co-dependence of  the bar fraction on color and M$_{\mathrm{*}}/$M$_{\mathrm{h}}$.
As reported by previous studies (Nair \& Abraham 2010; Lee+12), we detect a strong
dependence on color, with strong bars preferentially in red galaxies and weak bars
in blue systems. The contour colors indicate a stronger dependence on color than on the
stellar-to-halo mass ratio, but even at fixed $u - r$ color there is a clear dependence on
M$_{\mathrm{*}}/$M$_{\mathrm{h}}$, especially for values of
M$_{\mathrm{*}}/$M$_{\mathrm{h}}\geq-1.7$ with an increase of $f_{\mathrm{bar}}$ for
increasing M$_{\mathrm{*}}/$M$_{\mathrm{h}}$ at any $u-r$ value.

In Cervantes Sodi et al. (2013) we studied the dependence of the bar fraction
on the spin parameter $\lambda_{d}$ using the same galaxy sample employed here. 
To estimate the $\lambda_{d}$ we used a simple model introduced by
Hernandez \& Cervantes-Sodi (2006) with which we are able to give a first
order estimate of the spin in terms of the disk scalelength $R_{d}$, the circular
velocity $V$, the mass of the stellar disk $M_{d}$ and the disk mass fraction
$f_{d}$. For a detailed description of the model we refer the readers to
Cervantes Sodi et al. (2013). As a result, we found that the bar fraction strongly
depends on the spin, with the $f_{\mathrm{bar}}$ maximum at low to intermediate $\lambda_{d}$
values for the case of strong bars, while the maximum for weak bars is
at high $\lambda_{d}$. With $\lambda_{d}$ being proportional to the stellar-to-halo mass
ratio in our estimate, it is interesting to look if the bar fraction shows a dependence on
the spin, even at fixed M$_{\mathrm{*}}/$M$_{\mathrm{h}}$. The only difference between
present estimate for $\lambda_{d}$ with the one in Cervantes-Sodi et al. (2013) is that we are using 
M$_{\mathrm{*}}/$M$_{\mathrm{h}}$ from the group catalog, instead of the estimate by
Gnedin et al. (2007).

Figure~\ref{maps} bottom panels
show $f_{\mathrm{bar}}$ in the $\lambda_{d}$ vs. M$_{\mathrm{*}}/$M$_{\mathrm{h}}$
plane. It is evident that the dependence is stronger with M$_{\mathrm{*}}/$M$_{\mathrm{h}}$
than with $\lambda_{d}$, but still at fixed M$_{\mathrm{*}}/$M$_{\mathrm{h}}$ there is a variation
of the bar fraction with the spin parameter.
 For the case of strong bars (middle panel) with low
to intermediate values of M$_{\mathrm{*}}/$M$_{\mathrm{h}}$,
at fixed M$_{\mathrm{*}}/$M$_{\mathrm{h}}$ the bar fraction increases
for decreasing $\lambda_{d}$, while strong bars with heavy stellar disks
appear to be more frequently found on high spinning galaxies.
As discussed in Paper I, we expect galaxies with low spin values to be more
prone to develop bar instabilities due to their self-gravitation, while high
spinning galaxies are more extended with sparse disks that suppress and/or
damp global instabilities. This naturally explains the
behavior found in light disks, but disks with high stellar-to-halo mass ratios are
already massive enough to develop bar instabilities. For these systems with heavy disks 
in relation to their halos, the increase of $\lambda_{d}$ prevents them for becoming supported
by random motions instead of ordered rotation which in turn increases the formation/growth 
of bars, hence the increase of $f_{\mathrm{bar}}$ with increasing $\lambda_{d}$.
Weak bars (Figure~\ref{maps}, bottom right panel) have their maximum occurrence at low
M$_{\mathrm{*}}/$M$_{\mathrm{h}}$ ratios and high $\lambda_{d}$ values,
in agreement with recent numerical experiments (DeBuhr et al. 2012)
where weak bars appear only in marginally stable systems.

\begin{figure}

\includegraphics[width=.4\textwidth]{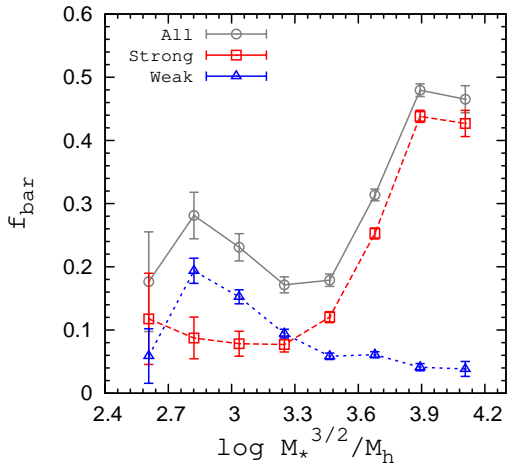} 

\caption[ ]{Fraction of barred galaxies $f_{\mathrm{bar}}$ as a function of
M$_{\mathrm{*}}^{3/2}/$M$_{\mathrm{h}}$ ratio.
\\(A color version of this figure is available in the online journal.)}\label{1.5power}
\end{figure}

\begin{figure*}[]
  \begin{center}
    \includegraphics [width=0.7\hsize]{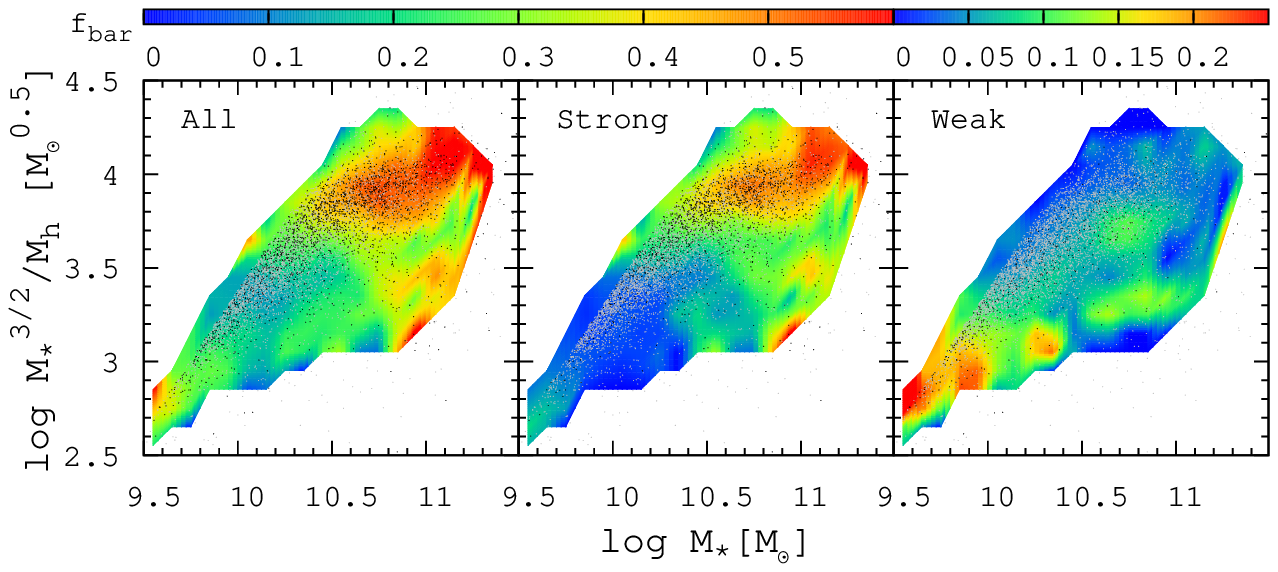}\\
    \includegraphics [width=0.7\hsize]{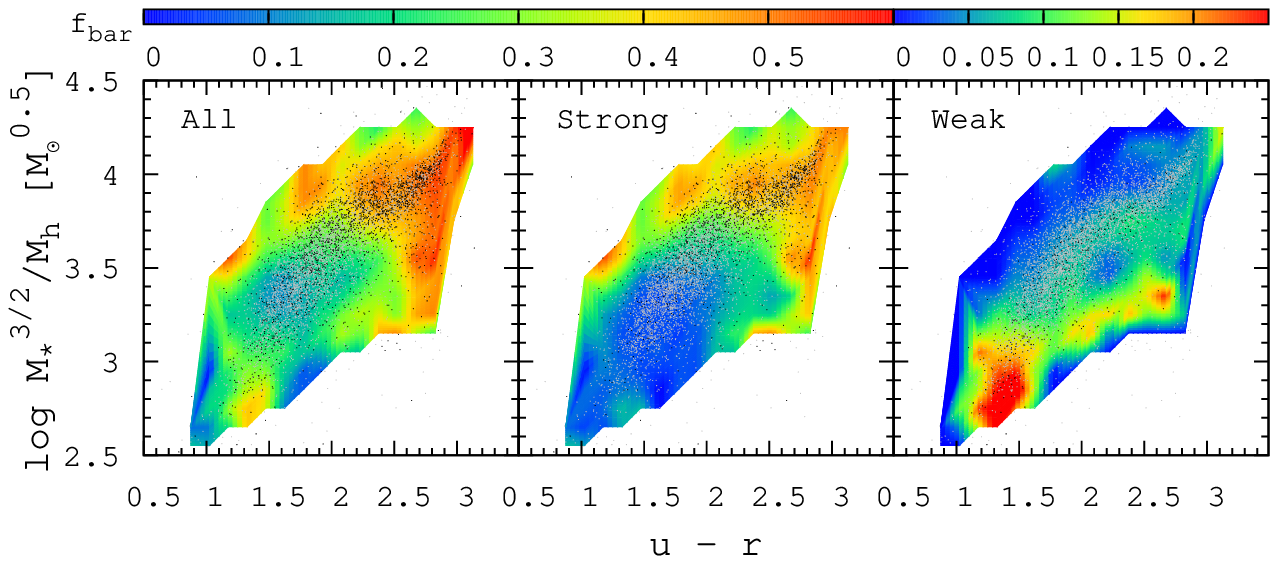} \\
    \includegraphics [width=0.7\hsize]{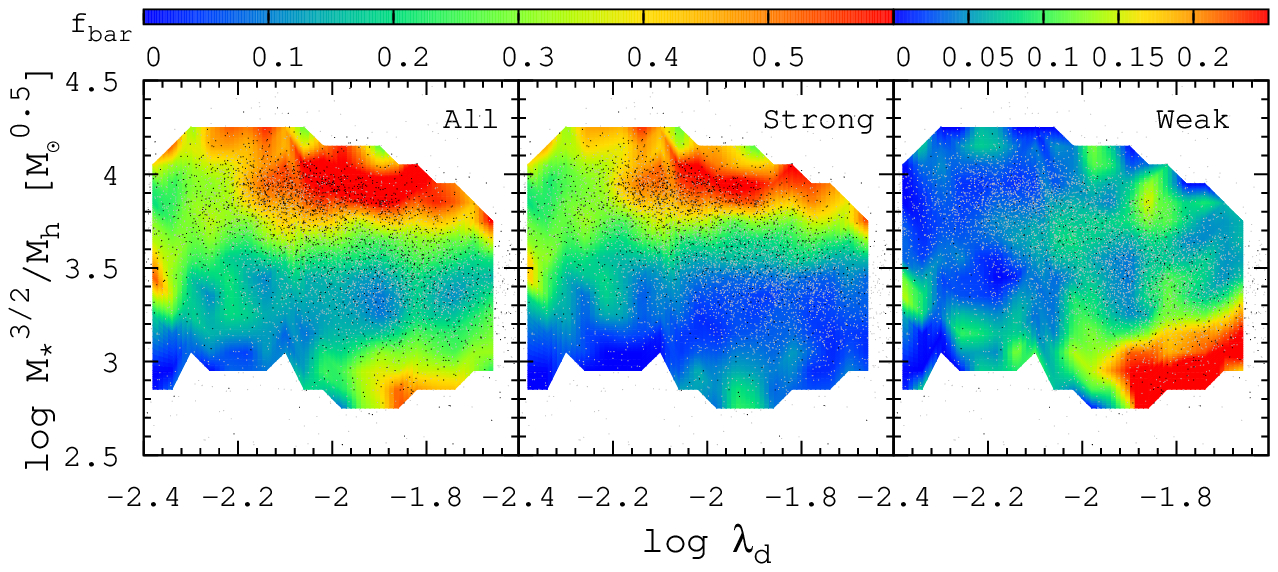}   
  
    \caption{Bar fraction $f_{\mathrm{bar}}$ isocontours in the M$_{\mathrm{h}}$ vs. M$_{\mathrm{*}}$ 
    (\textit{top panels}), $u-r$ vs. M$_{\mathrm{*}}/$M$_{\mathrm{h}}$ (\textit{middle panels}),
    and $\lambda_{d}$ vs M$_{\mathrm{*}}/$M$_{\mathrm{h}}$ (\textit{bottom panels})
    spaces. Left column correspond to strong plus
    weak bars, middle to strong bars and right to weak bars. The range and coding for $f_{bar}$ is
    the same for
    the full barred sample (strong plus weak bars) and for the limited sample of strong bars, with
    $0 \leq  f_{\mathrm{bar}} \leq 0.6$, while for the case of weak bars is restricted to
    $0 \leq  f_{\mathrm{bar}} \leq 0.25$. Gray dots represent unbarred
    galaxies, black dots represent barred ones.
    \\(A color version of this figure is available in the online journal.)}
  \label{maps2}
  \end{center}
\end{figure*}

The purpose of plotting a line with slope 1.5 in the log M$_{\mathrm{h}}$ vs. 
log M$_{\mathrm{*}}$ plane in Figure~\ref{maps} first panel is to make evident that moving perpendicular to this line
we get the most dramatic change in $f_{\mathrm{bar}}$, which is also shown in Figure~\ref{1.5power},
with a strong increase of $f_{\mathrm{bar}}$ with increasing M$_{\mathrm{*}}^{3/2}/$M$_{\mathrm{h}}$,
especially for the case of strong bars with log M$_{\mathrm{*}}^{3/2}/$M$_{\mathrm{h}}\gtrsim 3.3$,
where the increase of the bar fraction is more pronounced than the
one shown in Figure~\ref{Fractions}(c).

 A plausible explanation for the strong dependence of the
bar fraction on this ratio might arise from the stability criterion proposed by  Efstathiou et al. (1982). Using a
set of $N$-body simulations, they found that their systems were stable against bar formation if

\begin{equation}
\label{ELN}
\epsilon_{c} \equiv \frac{V_{max}}{(GM_{d}/R_{d})^{1/2}} < 1.1,
\end{equation}

where $V_{max}$ is the maximum rotation curve velocity, and $M_{d}$ and $R_{d}$ refer to the
mass of the disk and the disk scalelength respectively. In this sense, $\epsilon_{c}$ is a measure of the
self-gravity of the disk, very similar to the stability criterion proposed by Ostriker \& Peebles (1973) in
terms of the ratio of kinetic energy of rotation to total gravitational energy. If we consider a dark matter
halo with an isothermal density profile responsible for establishing a rigorously flat rotation curve
along the disk, then  $V_{max} \sim  (M_{\mathrm{h}}/R_{\mathrm{h}})^{1/2}$, and if $R_{d}$
is independent of $M_{d}$, then $\epsilon_{c} \sim M_{\mathrm{h}}/M_{d}^{3/2}$. 
This simple analysis presents the same functional dependence we obtain from our observational
sample and tells us that it comes from the different density distributions of dark matter and stars, and
reflects the stabilizing effect the halo provides to the disk of stars against bar formation.

For completeness, we also explore the behavior of $f_{\mathrm{bar}}$ in two-dimensional planes,
choosing one of the axes to be M$_{\mathrm{*}}^{3/2}/$M$_{\mathrm{h}}$, to check if at fixed
M$_{\mathrm{*}}^{3/2}/$M$_{\mathrm{h}}$ we can still find a dependence of the bar fraction
on other physical parameters. Our result is presented in Figure~\ref{maps2}. We can see that in all cases
(stellar mass, color and spin), the color contours denoting the bar fraction are almost
entirely horizontal in the areas most densely populated by galaxies, implying that
the bar fraction at fixed M$_{\mathrm{*}}^{3/2}/$M$_{\mathrm{h}}$ is practically independent
of stellar mass, color, and spin where the bulk of the galaxy population resides.
This independence is particularly interesting for the case
of color, for which previous studies have always found a strong dependence, even after
fixing other quantities; i.e. Nair \& Abraham (2010) found that, at fixed color, the bar fraction
still presents a dependence on stellar mass and morphological type; Masters et al. (2012)
found a dependence of the bar fraction on color at fixed absolute magnitude, and fixing the
color they also found a dependence on the gas mass fraction; and Lee+12 found a dependence
on $u - r$ color at fixed absolute $r-$magnitude, concentration index, and velocity dispersion.
Our result using central galaxies indicates, contrary to previous findings, that at fixed 
M$_{\mathrm{*}}^{3/2}/$M$_{\mathrm{h}}$, the dependence of the bar fraction on color
becomes much weaker, with the only exception of extreme red galaxies with $u-r \geq 2.6$.
Our result stresses the primary role that this parameter plays on establishing the conditions
for the presence of bars in disk galaxies, giving observational support to theoretical works (i. e.,
Yurin \& Springel 2014) that conclude that after the Efstathiou et al. (1982) stability criterion,
the importance of other physical parameters appear to be of secondary importance.

\begin{figure}
 \includegraphics[width=0.4\textwidth]{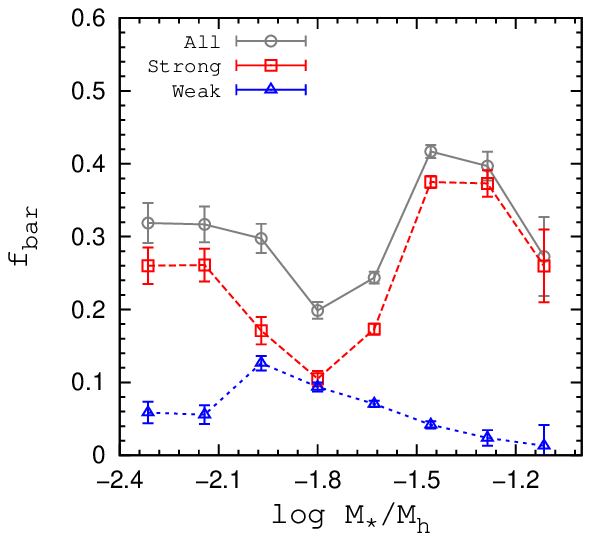}
\caption[]{Bar fraction $f_{\mathrm{bar}}$ as a function of the stellar-to-halo mass
ratio M$_{\mathrm{*}}/$M$_{\mathrm{h}}$ for the full sample (central plus satellite galaxies).
\\(A color version of this figure is available in the online journal.)}
 \label{centrals+satellites}
\end{figure}

\begin{figure}[]
\includegraphics[width=.475\textwidth]{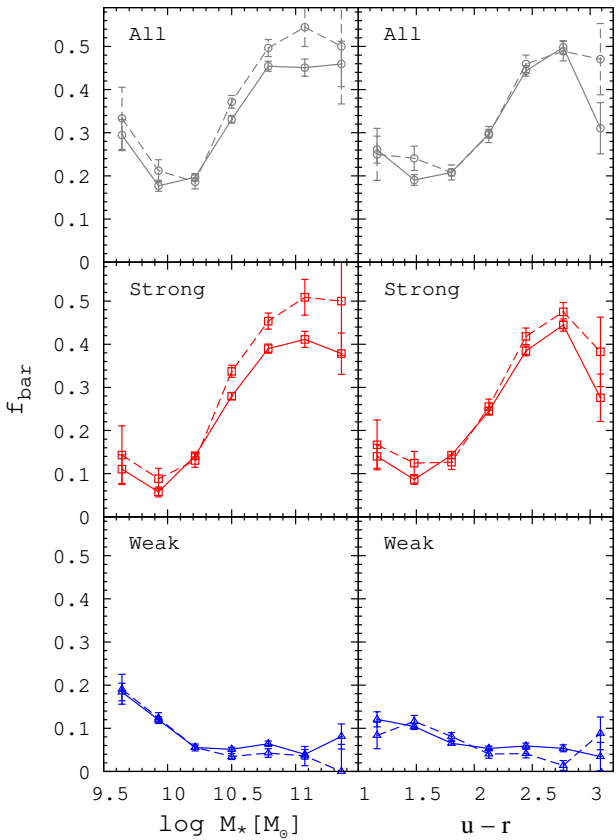} 

\caption[ ]{Dependence of bar fraction as a function of stellar mass (\textit{left column}),
and $u-r$ color (\textit{right column}) for all the galaxies in the sample (strong plus weak),
strong and weak bars. Solid line represents $f_{\mathrm{bar}}$ for central galaxies,
dashed line for satellite galaxies.
\\(A color version of this figure is available in the online journal.)}\label{Satellites}
\end{figure}

\subsection{Central plus satellite galaxies}

Before carrying on with our study using satellite galaxies, we consider it important to
explore what would be the result of studying the dependence of the bar fraction 
on the stellar-to-halo mass ratio
using the full sample of galaxies, centrals plus satellites, although in this case the
halo mass estimate for satellites does not correspond to their own dark matter halo, but
that of their central galaxy. If we compare this result of
Figure~\ref{centrals+satellites} with Figure~\ref{Fractions}c we
notice that the strong dependence is blurred in some degree, making less obvious the
dependence of  $f_{\mathrm{bar}}$ on M$_{\mathrm{*}}/$M$_{\mathrm{h}}$, helping
us to explain why this dependence has not been detected in previous works, where 
the distinction between centrals and satellites has not been taken into account
(Mart\'inez \& Muriel 2011).

The weak correlation between $f_{\mathrm{bar}}$ and the stellar-to-halo mass ratio for the full
sample is consistent with previous studies that have found no obvious dependence
of galaxy clustering on the presence of a bar (e.g. Li et al. 2009; Mart\'inez \& Muriel 2011; Lin et al. 2014).
Our analysis indicates that such dependence may be seen if the analysis was
restricted to central galaxies only.
This is certainly an interesting topic worthy of more work in future.

For galaxies in the satellite subsample, the stellar-to-halo mass ratio lacks importance,
given that the halo mass estimate corresponds to the parent halo of the most luminous
galaxy in the group, and we do not count with any estimate for the mass of the subhalo
hosting the satellite galaxy in question. In turn, we look at differences on the bar fraction
between central and satellite galaxies.

In a previous work using an extended sample from which our present sample is a subset, 
Lin et al. (2014) studied the environment of barred galaxies including early- and late-type
galaxies. They found early-type barred galaxies to be more strongly clustered on
 scales from a few 100 kpc to 1 Mpc when compared to early-type galaxies without bars.
Furthermore, they reported that for early-type galaxies, the fraction of central galaxies is
smaller if they host a bar, which indicates that the likelihood for early-type galaxies
to host a bar depends on the location within the dark matter halos. This result goes in the
same line as the one reported by Skibba et al. (2012), that shows a significant
environmental correlation of barred, bulge-dominated galaxies on the same
scales. Instead, when exploring the correlation function for the late-type galaxies of their
sample, Lin et al. (2014) found that the correlation function does not show any dependence
on the presence of bars, and that the ratio of central-to-satellite systems for barred and unbarred
late-type galaxies was very similar, about $\sim 27\%$.  In the present work, we look at differences of $f_{\mathrm{bar}}$
at fixed stellar mass and color for satellite and central galaxies in our sample, that corresponds to
the late-type subsample of Lin et al. (2014).

The left column of Figure~\ref{Satellites} shows $f_{\mathrm{bar}}$
as a function of stellar mass for central (solid line) and satellite (dashed line) galaxies.
For high mass galaxies with log M$_{\mathrm{*}}/$M$_{\mathrm{\odot}}\gtrsim10.5$
at a given stellar mass, the bar fraction of satellite galaxies is higher than the one for
centrals, with this effect seen only for the case of strong bars.
If we instead look at the bar fraction as a function
of $u-r$ color (same figure right column), the difference between satellites and centrals
vanishes. This can be explained with the satellite population being on average
redder at fixed stellar mass than the subsample of central galaxies, this due to the
harsh group environment where they experience gas stripping (Skibba 2009; Kimm et al. 2009;
Weinmann et al. 2010; Zhang et al. 2013), especially for low- to intermediate-mass galaxies. As they lose their
gas, their star formation shuts off and become red, which increases the likelihood for hosting
bars, as bars are more common in red galaxies with low gas content (Masters et al. 2012; Lee+12).
In this way the group environment is not directly responsible for an increase on the likelihood of
galaxies hosting bars, but indirectly by the dependence of $f_{\mathrm{bar}}$ on color.

Our result might help to explain why some authors find effects of the group/cluster
environment on the bar fraction (Andersen 1996), even at fixed luminosity (M\'endez-Abreu et al. 2012),
while others do not (van den Bergh 2002; Giordano et al. 2011; Mart\'inez \& Muriel 2011),
demonstrating that is important not only to control luminosity or stellar mass when comparing galaxies in different 
environments, but also color (see also Lee+12; Lin et el. 2014).

\section{Conclusions}

Using a volume-limited sample of galaxies with visually identified bars by Lee+12 and the Yang et al. (2007)
group catalog to discriminate between central and satellite galaxies and to estimate masses of
the parent halos of the groups, we investigated the dependence of the bar fraction on the stellar-to-halo 
mass ratio, finding that for central galaxies $f_{\mathrm{bar}}$ increases for increasing
M$_{\mathrm{*}}/$M$_{\mathrm{h}}$, even at fixed stellar mass or color. This result is in the same line with
early  (Ostriker \& Peebles 1973; Efstathiou et al. 1982) and recent (DeBuhr et al. 2012;
Yurin \& Springel 2014) theoretical works pointing out the stabilizing effect of dark matter halos on stellar disks against
bar formation.

Exploring the bar fraction in the  log M$_{\mathrm{h}}$ vs. log M$_{\mathrm{*}}$ plane,  
we find that the change of $f_{\mathrm{bar}}$
is the strongest if we consider a relation with the form
$f_{\mathrm{bar}}=f_{\mathrm{bar}}$(M$_{\mathrm{*}}^{\alpha}/$M$_{\mathrm{h}}$)
with $\alpha=1.5$. Furthermore, once M$_{\mathrm{*}}^{3/2}/$M$_{\mathrm{h}}$ is fixed,
the dependence of the bar fraction on stellar mass, color, and spin becomes very weak, with
the only exception of galaxies with extreme red colors.

By comparing the bar fraction for central and satellite systems at fixed stellar mass we
find that $f_{\mathrm{bar}}$ is higher for satellites with
log M$_{\mathrm{*}}/$M$_{\mathrm{\odot}}\gtrsim10.5$, but this difference vanishes when  we compare
the bar fraction at fixed $u-r$ color. We interpret this as follows: the bar fraction of satellites is higher
than that of centrals at given stellar mass (or luminosity) because satellites are redder than centrals on average at
this mass range. With $f_{\mathrm{bar}}$ being a strong function of color,
the difference of the bar fraction between satellites and centrals is not directly due to the
local environment, but indirectly through gas stripping suffered by the satellite population when
entering the group/cluster environment, with the subsequent reddening due to quenched star
formation. This, in turn, is what enhances the likelihood for galaxies to host bars, given that they
are more commonly found in red, gas-poor galaxies.

\acknowledgments
The authors thank the anonymous referee for useful comments that helped
to improve the quality of the paper and clarify our results.
B.C.S. thanks Ramin A. Skibba for valuable discussions about our results.
C.L. acknowledges the support of  the NSFC
(Grant Nos. 11173045, 11233005, 11325314, 11320101002)
and the Strategic Priority Research Program
``The Emergence of Cosmological Structures'' of CAS (grant No.
XDB09000000).
    Funding for the SDSS and SDSS-II has been provided by the Alfred P. Sloan Foundation,
    the Participating Institutions, the National Science Foundation, the U.S. Department of
    Energy, the National Aeronautics and Space Administration, the Japanese
    Monbukagakusho, the Max Planck Society, and the Higher Education Funding Council
    for England. The SDSS Web Site is http://www.sdss.org/. The SDSS is managed by the
    Astrophysical Research Consortium for the Participating Institutions. The Participating
    Institutions are the American Museum of Natural History, Astrophysical Institute Potsdam,
    University of Basel, University of Cambridge, Case Western Reserve University,
    University of Chicago, Drexel University, Fermilab, the Institute for Advanced Study, the
    Japan Participation Group, Johns Hopkins University, the Joint Institute for Nuclear Astrophysics,
    the Kavli Institute for Particle Astrophysics and Cosmology, the Korean Scientist Group,
    the Chinese Academy of Sciences (LAMOST), Los Alamos National Laboratory,
    the Max-Planck-Institute for Astronomy (MPIA), the Max-Planck-Institute for Astrophysics (MPA),
    New Mexico State University, Ohio State University, University of Pittsburgh,
    University of Portsmouth, Princeton University, the United States Naval Observatory,
    and the University of Washington.

\end{document}